\documentclass{aa}

\usepackage{graphicx}
\usepackage{txfonts}
\usepackage{siunitx}
\usepackage{caption}
\usepackage{subcaption}
\usepackage{natbib}
\bibpunct{(}{)}{;}{a}{}{,}
\usepackage[colorlinks=true,allcolors=blue]{hyperref}
\usepackage{orcidlink}

\begin{document}
    \title{Detection of colour variations from gravitational microlensing observations in the quadruple quasar HE0435-1223:\\Implications for the accretion disk}
    \titlerunning{Detection of colour variations from microlensing in HE0435-1223}
    \author{C. Sorgenfrei \and R. W. Schmidt \and J. Wambsganss\orcidlink{0000-0001-5055-7390}}
    \institute{Astronomisches Rechen-Institut, Zentrum für Astronomie der Universität Heidelberg,  Mönchhofstrasse 12-14, 69120 Heidelberg, Germany. \email{c.sorgenfrei@stud.uni-heidelberg.de}}
    \abstract{}
    {We present monitoring observations of quasar microlensing in the quadruple quasar HE0435-1223. The microlensing-induced light curves of the quasar images are chromatic, i.e. they depend on the applied filter band. Comparison with microlensing simulations allows us to infer properties of the accretion disk.}
    {We determine the $R$ and $V$ band light curves of the four images of HE0435-1223 from 79 and 80 epochs respectively, taken from 2014 to 2024 at the Las Cumbres Observatory using difference imaging analysis. We consider difference light curves to remove the intrinsic quasar variability. This reveals a prominent long-term chromatic microlensing event in image B. We use microlensing light curve simulations with both Gaussian and standard thin accretion disk brightness profiles to analyse this signal.}
    {The particularly strong signal observed in image B of HE0435-1223 makes it possible to detect a size ratio of the accretion disk in the $R$ band compared to the $V$ band of $1.24^{+0.08}_{-0.20}$ and $1.42^{+0.11}_{-0.22}$ for the Gaussian and the thin disk model, respectively. These values are in agreement with standard thin disk theory. For the absolute size we find large disk half-light radii of around $0.7$ to $1.0$ Einstein radii with an uncertainty of about $\SI{0.6}{dex}$ (depending on the filter bands and the models). Finally, our calculations show that image B undergoes caustic crossings about once per year.}{}
    \maketitle
    
\nolinenumbers 

\section{Introduction}
\label{sec:intro}

In the current picture of the structure of quasars, matter is accreted onto super-massive black holes in the centres of galaxies. Over the past decades the field has become quite mature \citep[see e.g.][]{Antonucci_1993,Frank_2002,Padovani_2017}, but obtaining direct constraints on properties of the accretion disk such as the actual size, the brightness profile or the temperature profile remains challenging. The widely used thin accretion disk model by \citet{Shakura_1973} predicts a power-law temperature profile of the accretion disk following
\begin{equation}
\label{equ:tempprofil}
    T(r) \propto r^{-\beta}\text{~~with~~}\beta=3/4,
\end{equation}
for large enough radii $r\gg r_{\text{in}}$ outside the inner edge of the disk \citep[we ignore here the inner cut-off term as well as different exponents, see e.g.][]{Abramowicz_1988,Mediavilla_2015,Vernardos_2024}.
While there are important constraints from reverberation mapping \citep[e.g.][and references therein]{Horne_2021}, direct observational tests of the accretion disk are difficult, since we typically cannot resolve the central quasar regions. However, quasar microlensing offers a way to `zoom' in.

Strong gravitational lensing creates multiple images of a quasar located (from the observers point of view) behind a massive galaxy acting as gravitational lens. The first discovery of such a lensed quasar was the doubly imaged quasar Q0957+561 \citep{Walsh_1979} and over 230 of these objects are confirmed today \citep{Ducourant_2018}.
By observing light curves (brightness as a function of time) of the multiple images and correcting for the time delays (due to the different light travel paths) between the images, two types of brightness variations can be distinguished: (1) Correlated brightness variations that occur in all images are intrinsic to the quasar. (2) Brightness variations only occurring in individual images, thus not originating from the quasar itself. 
These uncorrelated brightness variations are caused by compact objects in the light path, such as stars in the lensing galaxy in the vicinity of the line of sight to a quasar image. The compact objects exert an additional lensing of the quasar images with deflection angles on the scale of micro-arcseconds. Through the relative motion of quasar, lens and observer, a time-varying magnification acts on the quasar images on time scales of weeks, months and even longer, which leads to these uncorrelated brightness variations \citep[see][]{Chang_1979,Schmidt_2010,Vernardos_2024}. 
This so-called quasar microlensing effect was first detected in the quadruply imaged quasar Q2237+0305 \citep{Irwin_1989,Corrigan_1991}, a source which was discovered by \citet{Huchra_1985} and is also known as the `Einstein Cross'.

Microlensing signals can be used to test model predictions for the structure of quasars accretion disks, such as their size \citep{Kochanek_2004,Poindexter_2010,Morgan_2010,Morgan_2018,Cornachione_2020b}. The effect of microlensing in the plane of the quasar (source plane perpendicular to the line of sight) can be described by a pattern of varying magnification with characteristic caustic lines (lines of formal infinite magnification) produced by the combined effect of the compact objects
\citep[e.g.][]{Kayser_1986,Wambsganss_1990}. The microlensing signal depends on the size of the source as it is moving through the caustic pattern and averaging the magnifications over its size according to the brightness profile. 
Importantly, this means that also the slope of the temperature profile of the accretion disk is accessible \citep{Wambsganss_1991,Anguita_2008,Eigenbrod_2008,Poindexter_2008,Mosquera_2009,Cornachione_2020a}, since observing microlensing events in different filters corresponds to microlensing of different source sizes, where the disk size depends on the filter wavelength according to
\begin{equation}
\label{equ:lambdasize}
    r \propto \lambda^{4/3},
\end{equation}
assuming that at all radii the accretion disk radiates as a black body with temperature according to Eq. \ref{equ:tempprofil}.
 
In a previous study \citep{Sorgenfrei_2024} we described our method to obtain long-term light curves of multiply imaged quasars using point spread function (PSF) photometry on images generated with difference imaging analysis (DIA), where we used \textit{Gaia} proper motion data \citep{GAIA_2016,GAIA_2023} to improve image alignment and quasar image positions. In that study, we applied this method to observations in the $R$ and $V$ band taken at the Las Cumbres Observatory \citep[LCO\footnote{\url{https://lco.global/}},][]{Brown_2013} of the three lensed quasars HE1104-1805, HE2149-2745 and Q2237+0305 and determined their light curves from 2014 to 2022. We have since updated their light curves to include data until March 2024 and additionally reduced LCO data of another lensed quasar, HE0435-1223, leading to light curves of each image of the four quasars in both bands over 10 years.\footnote{The updated (and new) light curves of the four quasars (and more in the future) are available at \citet{GAVO_2025}, including data until March 2024. For HE1104-1805 they now consist of 124 $R$ and 126 $V$ band, for HE2149-2745 of 175 $R$ and 146 $V$ band and for Q2237+0305 of 115 $R$ and 114 $V$ epochs, which is overall a plus of $\sim39\%$ with respect to \citet{Sorgenfrei_2024}.} 
In this paper we focus on HE0435-1223, utilizing the light curve data in both bands with respect to determining constraints on the accretion disk temperature profile.

\begin{figure*}
    \centering
    \includegraphics[width=0.97\textwidth]{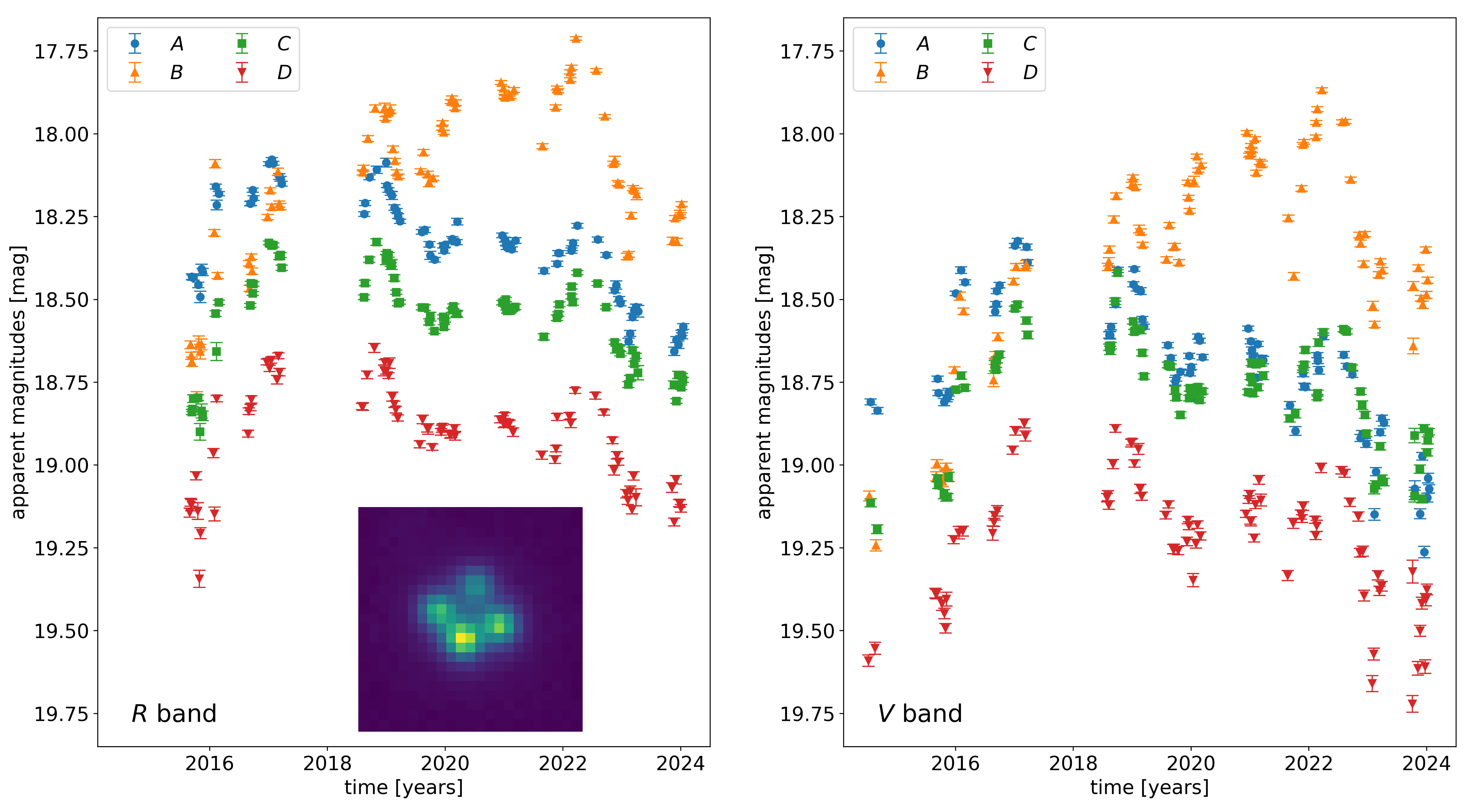}
    \caption{Light curves of HE0435-1223. This figure shows the time delay corrected light curves of the four images of HE0435-1223 in the $R$ band (left) and $V$ band (right) with the same time and magnitude ranges, with $1\sigma$ magnitude errors. Inside the left panel we have inserted the DIA reference image of HE0435-1223 in the $R$ band (linear in flux) as an example. The image size is $\SI{9}{arcsec}\times\SI{9}{arcsec}$, quasar image A is on the right and images B, C and D follow clockwise. Here, image B appears to be brightest, since the $R$ band reference image is combined mostly from images of the year 2022.}
    \label{fig:lightcurves}
\end{figure*}

HE0435-1223 is a quadruply imaged quasar discovered by \citet{Wisotzki_2002}. The four images are observed in a `cross' configuration similar to the Einstein Cross with comparatively wide separations (see Table \ref{tab:HSTdistances}). The quasar is located at a redshift $z_\text{s}=1.693$ \citep{Sluse_2012}. It is lensed by a foreground galaxy at $z_\text{L}=0.454$ \citep{Eigenbrod_2006}, which can be seen in \textit{Hubble} Space Telescope (HST) images \citep{Falco_2001} but is not very prominent in our observations (see inset on the left panel in Fig. \ref{fig:lightcurves}). The time delays between the images as determined by COSMOGRAIL are $\Delta t_{AB}=\SI[separate-uncertainty=true]{-9.0\pm0.8}{days}$, $\Delta t_{AC}=-0.8^{+0.8}_{-0.7}\,\SI{}{days}$ and $\Delta t_{AD}=\SI[separate-uncertainty=true]{-13.8\pm0.8}{days}$, where image A is leading \citep{Millon_2020}.
\begin{table}
    \sisetup{separate-uncertainty=false}
    \caption{HE0435-1223 image position separations.}
    \centering
    \begin{tabular}{ c c c c }
        \hline\hline
        separation $[\SI{}{arcsec}]$ & B$-$A & C$-$A & D$-$A\\
        \hline
        right ascension $\Delta\alpha$ & $\SI{-1.476\pm0.003}{}$ & $\SI{-2.467\pm0.003}{}$ & $\SI{-0.939\pm0.003}{}$\\
        declination $\Delta\delta$ & $\hphantom{-}\SI{0.553\pm0.003}{}$ & $\SI{-0.603\pm0.005}{}$ & $\SI{-1.614\pm0.003}{}$\\
        \hline
    \end{tabular}
    \label{tab:HSTdistances}
    \sisetup{separate-uncertainty=true}
    \tablefoot{The image position separations in arcseconds with uncertainties (quasar images B, C and D with respect to A) are taken from the CASTLES webpage \url{https://lweb.cfa.harvard.edu/castles/} (by C.S. Kochanek, E.E. Falco, C. Impey, J. Lehar, B. McLeod, H.-W. Rix), that uses \textit{Hubble} Space Telescope (HST) data \citep{Falco_2001}.}
\end{table}

We summarize the main steps of the data reduction with the resulting light curves in Sect. \ref{sec:data} and present the microlensing signal we found. In Sect. \ref{sec:MLsim} we describe a set of microlensing simulations to investigate this signal. We present and discuss our results in Sect. \ref{sec:results} and conclude in Sect. \ref{sec:conclusion}.

\section{From LCO data to microlensing signal}
\label{sec:data}

In order to investigate quasar microlensing, we use $R$ and $V$ band data taken at LCO, a global network of robotic telescopes. Targeting multiple lensed quasars, data were acquired since 2014 by \SI{1}{m} telescopes at five locations (mainly Cerro Tololo, Chile\footnote{For the final light curves of HE0435-1223, only data taken at Cerro Tololo, Chile was used, since a low number of observations from other locations were excluded at various steps of the data reduction.}; but also: Sutherland, South Africa; Siding Springs, Australia; McDonald, USA; Teide, Spain), with a total of 677 and 697 observations of HE0435-1223 between 24 July 2014 and 17 January 2024 (many during a single night) in the $R$ and $V$ bands, respectively. Since the method to reduce the data and determine the light curves is the same as in \citet{Sorgenfrei_2024}, we only summarize the main steps in the following subsection and refer to our previous paper for the implementation\footnote{\url{https://github.com/sorgenfrei-c95/qsoMLdiffcurves}} and the details.

\subsection{Data reduction}
Each observation is aligned to a reference image using the alignment routine of the \texttt{ISIS} software\footnote{\url{http://www2.iap.fr/users/alard/package.html}} by \citet{Alard_2000}. We modified it to include \textit{Gaia} proper motion data to improve the alignment by correcting the determined star positions in the reference image by their proper motion and the time between observation and reference image. For HE0435-1223, the median dispersion of star position deviations between observation and reference image after image alignment \citep[see Eq. 2 in][]{Sorgenfrei_2024} improved from $\left<\sigma\right>\!\sim\!\SI{59}{mas}$ before to $\sim\!\SI{43}{mas}$ after this \textit{Gaia} correction. Afterwards images from one night were combined leaving us with 79 epochs in the $R$ band as well as 80 in the $V$ band. We additionally produced a high S/N reference image from a few low-seeing observations for the DIA (see inset on the left panel of Fig. \ref{fig:lightcurves}).

In preparation for applying the DIA method, we use PSF photometry to determine the quasar image A positions in the combined images and the PSFs as was done in \citet{Giannini_2017a}. For that we use \texttt{GALFIT} \citep[][version 2.0.3]{Peng_2002}, but modified to fit a multiple quasar image model consisting of copies of the PSF (a fixed $30\times30$ pixel cutout of a star in the vicinity of the quasar) with fixed relative positions determined from HST images; for HE0435-1223 the fixed separations are given in Table \ref{tab:HSTdistances}. Applying this model to each of our combined images, \texttt{GALFIT} finds the best value for the position of quasar image A.

Next, DIA is applied to all images, to obtain the difference between each combined image and the reference image. In detail, for each combined image, the reference image is convolved with a continuous kernel function estimated from $5\times5$ stamp stars that are scattered over the whole image to correct for seeing differences. This is done using the so-called \texttt{hotpants} software\footnote{\url{https://github.com/acbecker/hotpants}} by \citet{Becker_2015}, an implementation of the algorithm from \citet{Alard_1998} and \citet{Alard_2000}. The resulting difference images contain only brightness variations with respect to the reference image. Therefore light from the lens galaxy is removed and only brightness variations of the four quasar images remain at their positions.

Finally, the quasar light curves are extracted from the difference images by applying PSF photometry with the quasar image A position of our multiple quasar image model and PSF fixed from before, and then adding the resulting difference fluxes to the quasar image fluxes from PSF photometry of the reference image. The zero point of the apparent magnitude scale is determined by the brightness of several stars in the $R$ and $V$ band reference images relative to their apparent magnitudes determined from \textit{Gaia} $G$, $G_{bp}$ and $G_{rp}$ data \citep[see][]{Riello_2021}.

\subsection{Light curves of HE0435-1223}
The final light curves (apparent magnitudes over time) are shown in Fig. \ref{fig:lightcurves} and consist of 79 and 80 data points for all four images in the $R$ and the $V$ band, respectively. The light curves of images B, C and D are shifted in time by the delays given in Sect. \ref{sec:intro}. The intrinsic quasar variability is immediately visible in all images and both bands, e.g. the brightness peaks around 2019 and in 2022, as well as the general shape of the curves, except for the strong increase of image B compared to the other images. In the 10 years of observations the apparent magnitude of image B changes by up to $\sim\SI{1.0}{mag}$ in the $R$ band and by even $\sim\SI{1.3}{mag}$ in the $V$ band.

\subsection{Difference curves and microlensing signal}
\label{sec:diffcurves}
We interpret, that the additional variability of image B noticeable in Fig. \ref{fig:lightcurves} is due to microlensing as described in Sect. \ref{sec:intro}. In order to isolate this additional microlensing signal quantitatively we calculate difference curves between pairs of observed light curves. This removes the intrinsic quasar variability present in all images. Since the light curves of the individual images have to be corrected for time delays, interpolation between data points is required to calculate this difference. This  can sometimes be problematic for time delays for which the shifted light curve is moved into seasonal observation gaps of the other light curve. However, the time delays of HE0435-1223 are of the order of just a few days \citep[][see Sect. \ref{sec:intro}]{Millon_2020}, resulting in only slightly shifted light curves with a large overlap and no need for any uncertain interpolation across gaps of the data.

\begin{figure*}
    \centering
    \includegraphics[width=0.97\textwidth]{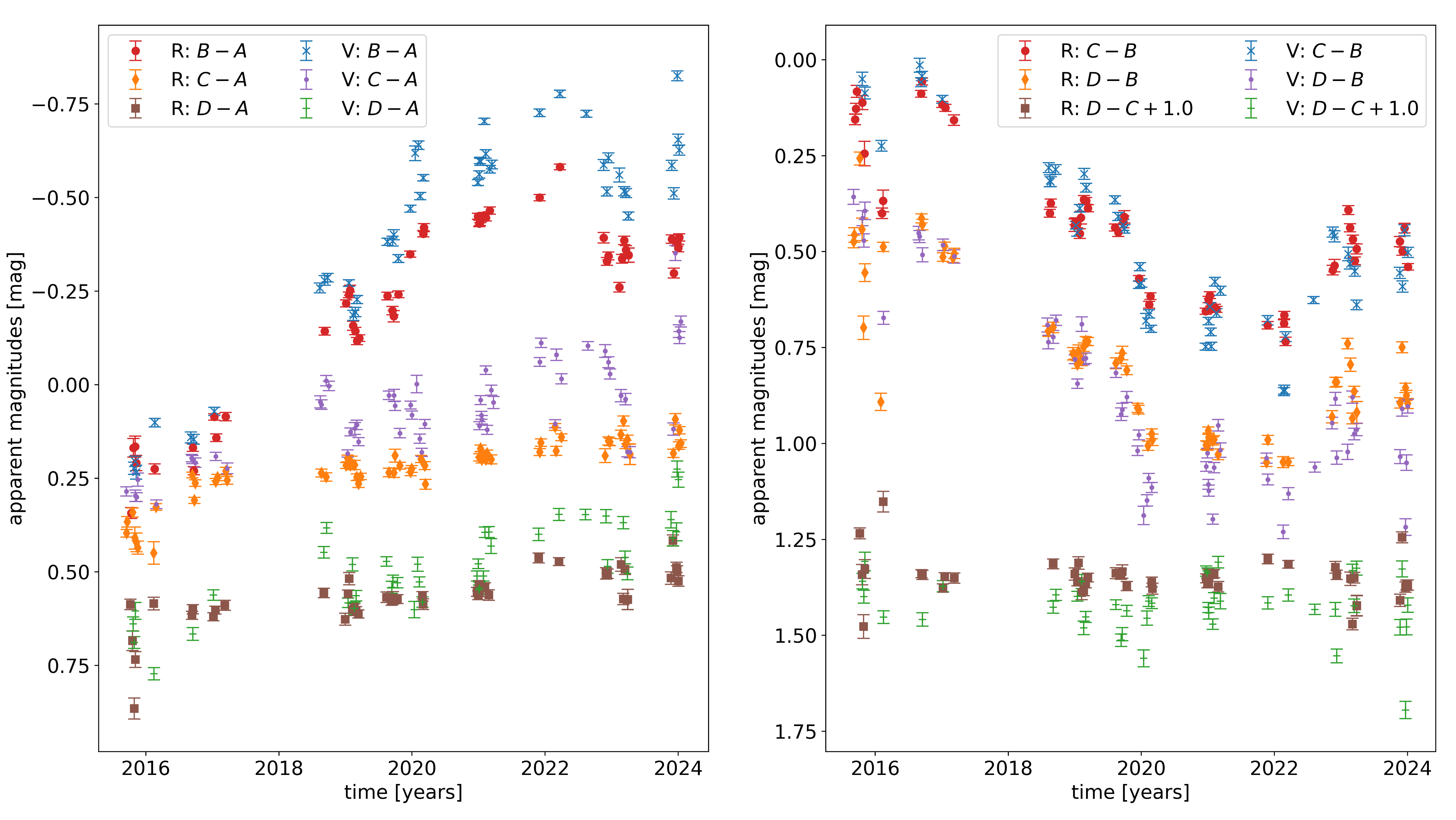}
    \caption{Difference curves of HE0435-1223. Shown are the six combinations of differences of the four light curves ($B-A$, $C-A$ and $D-A$ on the left side and $C-B$, $D-B$ and $D-C+\SI{1.0}{mag}$ on the right side) both in the $R$ band (red circles, orange diamonds and brown squares) and the $V$ band (blue crosses, violet dots and green minuses) with $1\sigma$-uncertainties.}
    \label{fig:diffcurves}
\end{figure*}

We use linear interpolation of image A to the time delay corrected epochs of images B, C and D, as well as B to C and D and finally C to D. Interpolation is only done between data points separated by $\leq\SI{30}{days}$. Errors of the interpolated light curves are estimated by Monte Carlo sampling assuming Gaussian errors of the data points. Then, the difference curves are simply the differences between a given reference light curve and the suitably interpolated light curve, e.g. $B-A := m_B(t+\Delta t_{AB})-m_{A\text{,\,interp.}}(t)$, with standard Gaussian errors. The resulting six difference curves are shown in Fig. \ref{fig:diffcurves}.

In all three combinations of difference light curves including image B the already-mentioned strong microlensing variation can be seen. A colour effect is also visible: the variation has a larger amplitude in the $V$ band than in the $R$ band. However, the three combinations using image A show an additional long term microlensing signal in the form of an approximately linear rise, which indicates that the $B-A$ curve includes the combination of microlensing signals from two lines of sight and is thus more difficult to analyse \citep[e.g.][]{Eigenbrod_2008,Anguita_2008}. Nevertheless, the difference between images C and D (bottom curve in the right hand panel of Fig. \ref{fig:diffcurves}) is mostly flat implying no (or little) microlensing in these images, therefore making the $C-B$ (or $D-B$) curve a suitable target for a microlensing analysis, with the microlensing signal coming chiefly from one line of sight. Consequently, we restricted our microlensing analysis to the signal in the (inverted) $B-C$ curve, thus avoiding the combinatoric explosion arising from a necessity of simulating microlensing in all images simultaneously \citep[e.g.][]{Kochanek_2004}. We repeated the following simulation (Sect. \ref{sec:MLsim}) and subsequent analysis (Sect. \ref{sec:results}) using the $B-D$ curve (which is affected more by noise, since image D is fainter). The results are consistent (with slightly larger uncertainties).

\section{Microlensing Simulations}
\label{sec:MLsim}

In order to quantitatively analyse the (chromatic) microlensing signal of image B in the difference curve $B-C$, we simulate microlensing light curves and compare them to our data, mainly following the method developed in \citet{Kochanek_2004} and used by numerous works to analyse microlensing in multiple lensed quasar systems afterwards \citep[e.g.][]{Anguita_2008,Morgan_2010,Morgan_2018,Cornachione_2020a,Cornachione_2020b}. 

\subsection{Magnification patterns}
\label{sec:magpat}
First, source-plane magnification patterns with the microlensing parameters of image B were generated using \texttt{Teralens}\footnote{\texttt{Teralens} is a microlensing code developed by \citet{Alpay_2019} for use on Graphics Processing Units (GPUs) available at \url{https://github.com/illuhad/teralens}. For reference we note that the shear direction in \texttt{Teralens} is rotated by $\pi/2$ when compared to \texttt{microlens} \citep{Wambsganss_1999}. We have added the output of the mean magnification to the default version of \texttt{Teralens} in order to be able to calculate magnitudes.} \citep{Alpay_2019}. It is a new implementation of the inverse ray shooting tree code from \citet{Wambsganss_1999}, fully parallelized for GPUs. 
We fix the total convergence and shear to $\kappa=0.539$ and $\gamma=0.602$ \citep{Schechter_2014} and vary the fraction of compact matter $\kappa_\star$ in steps of $0.1$ from $\kappa_\star/\kappa = 0.1$ to $1.0$ (the convergence is composed of contributions from compact and smooth matter $\kappa=\kappa_\star+\kappa_{\text{smooth}}$, as usual). We assume equal masses of $M=1M_{\odot}$ for all microlenses in the lens plane \citep{Lewis_1996} and produce ten statistically independent magnification patterns with sizes of $\SI{40}{\textit{R}_E}\times\SI{40}{\textit{R}_E}$ and a resolution of $\SI{200}{pixel/\textit{R}_E}$. Using a flat cosmology with $H_0=\SI{70.0}{kms^{-1}Mpc^{-1}}$, $\Omega_{\text{m,0}}=\SI{0.3}{}$, as well as the redshifts from Sect. \ref{sec:intro}, the source-plane Einstein radius (which scales with the square root of the true mean mass $\left<M\right>$ of the microlenses) corresponds to
\begin{equation}
\label{equ:einsteinradius}
    R_\text{E} = \sqrt{\frac{4\text{G}\left<M\right>}{\text{c}^2}\frac{D_\text{S} D_\text{LS}}{D_\text{L}}}\simeq 5.42\times10^{16}\sqrt{\left<M\right>\!/M_{\odot}}\;\SI{}{cm},
\end{equation}
where $D_\text{S}$, $D_\text{L}$ and $D_\text{LS}$ are the angular diameter distances from observer to source, observer to lens and lens to source.

As mentioned in Sect. \ref{sec:intro}, the apparent source size connected to the quasar's temperature profile influences the observed microlensing signal, which is the signal we are interested to find. Inserting the temperature profile (Eq. \ref{equ:tempprofil}) of a quasar from the thin disk model \citep{Shakura_1973} into Planck's law (assuming a small filter width) results in a radial surface brightness profile
\begin{equation}
\label{equ:brightnessprofil}
    B(r) \propto \left[\exp\left(\left(\frac{r}{r_s}\right)^{3/4}\right)-1\right]^{-1}
\end{equation}
with a characteristic scale radius $r_s$ (where the temperature $T(r_s)=hc/(k_B\lambda_0)$ matches the filter wavelength in the quasar rest frame  $\lambda_0$). However, often this profile is replaced with a Gaussian brightness profile
\begin{equation}
\label{equ:gaussprofil}
    B(r) \propto \exp\left(-\frac{r^2}{2R_s^2}\right)
\end{equation}
with a (different) scale radius $R_s$. This can be used instead, since it was shown by \citet{Mortonson_2005}, that primarily the overall size measured in half-light radii $r_{1/2}$ is important for the microlensing effect, rather than the detailed shape of the radial profiles. Integrating the profiles, assuming circular symmetry, results in half-light radii of $r_{1/2}\simeq2.44r_s$ for the Shakura-Sunyaev thin disk profile (Eq. \ref{equ:brightnessprofil}) and $r_{1/2}\simeq1.18R_s$ for the Gaussian disk (Eq. \ref{equ:gaussprofil}), both viewed face-on.

Nevertheless, we shall consider both types of profiles in this study. We produce circular symmetrical kernels (going out to $10$ scale radii or at most $\SI{10}{\textit{R}_E}$) as a model for the brightness of face-on quasar accretion disks. We therefore convolve our 10 ($\kappa_\star/\kappa$ varying) magnification patterns for image B with these kernels made from 80 different scale radii values for both the Shakura-Sunyaev thin disk and the Gaussian brightness profile, in total yielding $1600$ convolved maps. 

The 80 size values were chosen non-logarithmically over a large range resulting in an approximately even sampling of size ratios \citep[see][more on this in Sect. \ref{sec:results}]{Anguita_2008,Eigenbrod_2008}. We test the same size values as the scale radius parameter for both the thin disk $r_s$ and the Gaussian disk $R_s$, starting with 0.5, 1, 2, 3, 4, 6, 8, 10 pixels, continuing in steps of 5 pixels up to a size of 200 pixels, then in 10 pixel steps up to 400 pixels, in 20 pixel steps up to 600 pixels and finally in 50 pixel steps up to 800 pixels. Therefore, we test disks with scale radii as small as $\SI{0.0025}{\textit{R}_E}$ and as large as $\SI{4.0}{\textit{R}_E}$ corresponding to $r_{1/2}\simeq\SI{0.0029}{\textit{R}_E}$ or $\SI{0.0056}{\textit{R}_E}$ (around the pixel scale) up to $r_{1/2}\simeq\SI{4.71}{\textit{R}_E}$ or $\SI{8.96}{\textit{R}_E}$ for the Gaussian and thin disks half-light radii, respectively. In order to avoid edge effects for the largest disks from the convolution (using a Fast-Fourier-Transform which assumes periodical patterns), we discard the outer $\SI{5}{\textit{R}_E}$ on all sides of each convolved magnification map. This leaves us with convolved magnification patters of $\SI{30}{\textit{R}_E}\times\SI{30}{\textit{R}_E}$.

\subsection{Light curve fitting}
\label{sec:lcfit}
\begin{figure}
    \resizebox{\hsize}{!}{\includegraphics{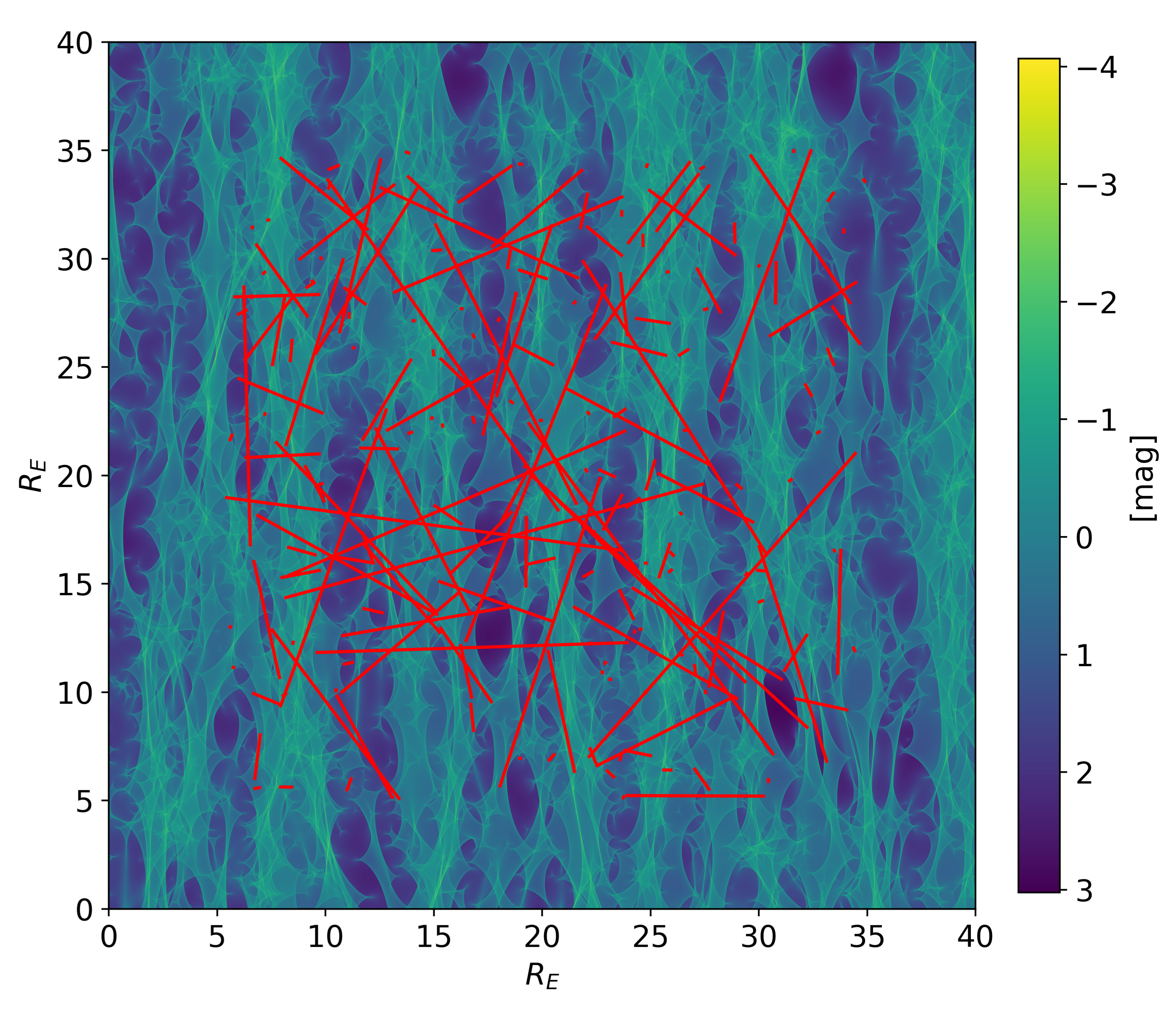}}
    \caption{Magnification pattern with example tracks. Shown are the first 200 tracks (in red) on the unconvolved magnification map of image B as produced by \texttt{Teralens} for $\kappa_\star/\kappa=0.8$.}
    \label{fig:mapwithtracks}
\end{figure}

In order to extract light curves from the 1600 convolved magnification maps, we construct $10^7$ tracks with track lengths corresponding to 10 years by drawing $10^7$ velocities sampled from a log-uniform distribution between $\SI{0.002}{\textit{R}_E/yr}$ and $\SI{2.0}{\textit{R}_E/yr}$ (in principle only limited by the size of our magnification patterns) with random directions and random $x$ and $y$ start point coordinates (inside the central $\SI{30}{\textit{R}_E}\times\SI{30}{\textit{R}_E}$ with redrawing of just the start points of those tracks, which leave that region). As an example, in Fig. \ref{fig:mapwithtracks} we show the unconvolved magnification map for $\kappa_\star/\kappa=0.8$ generated with \texttt{Teralens} together with 200 tracks, showcasing the range of track lengths (i.e. velocities), random directions, as well as the restriction to the central $\SI{30}{\textit{R}_E}\times\SI{30}{\textit{R}_E}$. 

For each compact object fraction $\kappa_\star/\kappa$, each disk size $r_s$ or $R_s$ and both disk models, we place all tracks on the corresponding map and interpolate the magnitudes of 500 equal steps along each track. The generated light curves are then compared to the $R$ and $V$ band $B-C$ difference curves from Sect. \ref{sec:diffcurves}, by fixing the start time of the simulated tracks to $t_0=\SI{2014.5}{yr}$ and then interpolating them to the epochs $t_i$ of the difference curves, thus extracting two simulated microlensing curves of image B at the same epochs of the $R$ and $V$ difference curves respectively. Since we assume that there is no microlensing along the line of sight of image C, we subtract a constant offset $\mu_C$ (which takes into account the mean magnifications from strong lensing and a possible non-zero, but constant microlensing signal in image C) from each simulated $\mu_B(t_i)$ curve. The offset $\mu_C$ is chosen such, that the difference of the mean magnitude values of simulated and measured difference curves is minimized. The resulting goodness-of-fit estimator $\chi^2$ is
\begin{equation}
\label{equ:chi2}
    \chi^2 = \sum_i\left[\frac{\mu_B(t_i)-\mu_C-(B-C)(t_i)}{\sigma_i}\right]^2,
\end{equation}
which is essentially Eqs. 6 and 7 of \citet{Kochanek_2004} applied to our case, with the $\sigma_i$ set to the magnitude errors of $B-C$ at epochs $t_i$, where we added $\SI{0.05}{mag}$ in quadrature to the $1\sigma$-errors of both light curves analogous to \citet{Kochanek_2004}. Subsequent works have incorporated unknown systematic uncertainties similarly \citep[e.g.][]{Eigenbrod_2008,Poindexter_2010,Morgan_2018}. This remains necessary for our simulations as well, because our microlensing model does not explain these remaining residuals \citep[see e.g.][for a recent discussion of possible implications]{Paic_2022}.

In all $10^8$ cases ($10^7$ tracks on 10 independent maps) and for both models, we determine the best-fitting disk size of all 80 tested sizes, that minimizes $\chi^2$ from Eq. \ref{equ:chi2}. This is done independently for $R$ and $V$ and since we want our simulated light curves to reproduce both the $R$ and $V$ band data, we add these best-size $\chi^2_R$ and $\chi^2_V$ values to a combined $\chi^2=\chi^2_R+\chi^2_V$. This results in a $\chi^2$-value with a best disk size in $R$ and $V$ for each of the $10^7\times10$ tracks and both disk models.

\begin{figure*}
    \centering
    \includegraphics[width=0.97\textwidth]{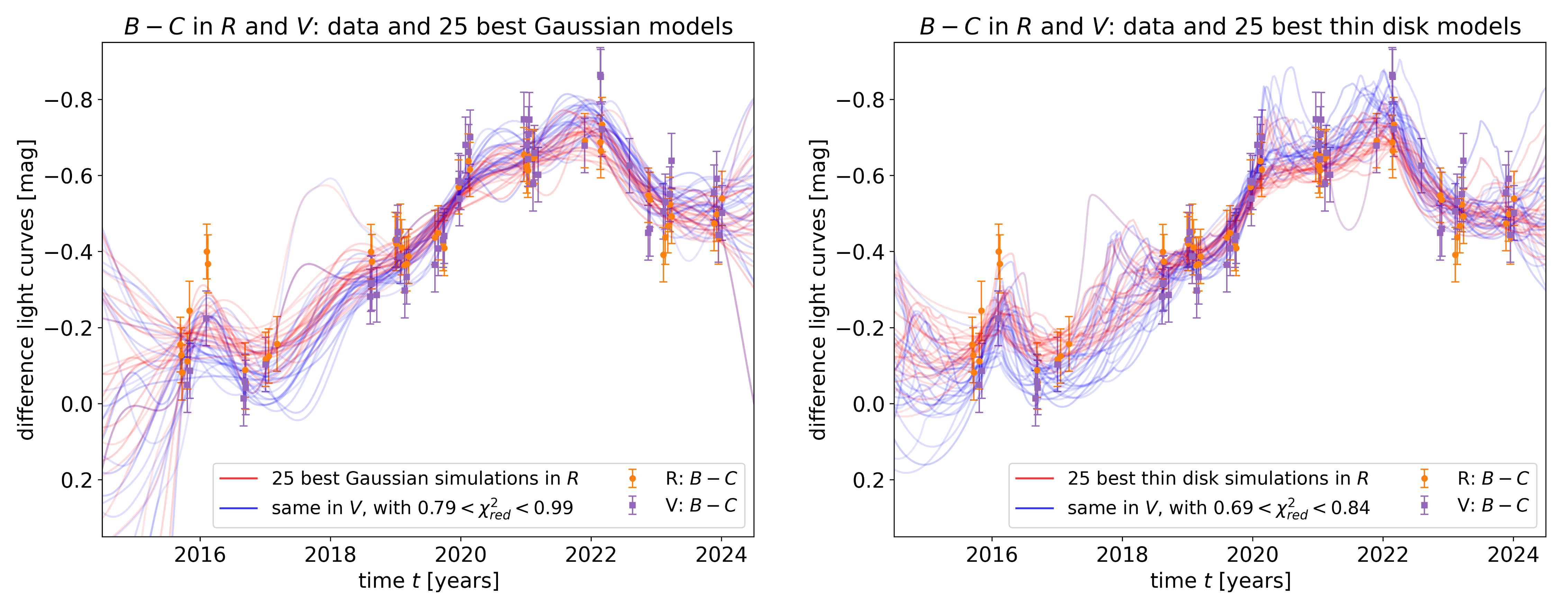}
    \caption{Best-fitting simulated light curves. We show here the best-fitting $R$ and $V$ simulated microlensing light curves (red and blue curves) from the 25 tracks with smallest $\chi^2=\chi^2_R+\chi^2_V$ for the Gaussian disk model (left) and the Shakura-Sunyaev thin disk model (right). The data points are the measured $R$ (orange circles) and $V$ (purple squares) $B-C$ difference curves (i.e. the two uppermost curves on the right side of Fig. \ref{fig:diffcurves} inverted and with adapted error bars as described in Sect. \ref{sec:lcfit}) as used for the $\chi^2$ calculation.}
    \label{fig:simcurves}
\end{figure*}

These $\chi^2$-values are then converted to `track probabilities' $P(\chi^2)$, the likelihood of each simulated $R$ and $V$ band light curve pair, using Eq. 10 from \citet{Kochanek_2004}.\footnote{Therefore we use $P(\chi^2)\propto\Gamma[N_{\text{dof}}/2-1,\chi^2/(2f_0^2)]$, which implies a rescaling of $\chi^2_f=\chi^2/f^2$ with $P(f)\propto f$ for $0\leq f\leq f_0$ \citep[see][]{Kochanek_2004}. For our combined $R$ and $V$ band $B-C$ data the number of degrees of freedom is $N_{\text{dof}}=98$. We set $f_0=\text{min}(3\chi_{\text{red}}/2)$, which is of order unity, independently for thin and Gaussian disk, chosen such that for both the minimum rescaled reduced $\chi^2_f/N_{\text{dof}}$ has an expectation value of one. This means that we can compare both disk models although their $\chi^2$-distributions differ slightly.}
We apply this procedure only to tracks for which $\chi^2_{\text{red}}\leq 5$ holds for the reduced $\chi^2_{\text{red}}=\chi^2/N_\text{dof}$, speeding up the computations by removing models with vanishing likelihood.
Finally, these results are collected in a `track library' including track number, $\kappa_\star/\kappa$, velocity, direction, the four best sizes and the two track probabilities. About $\SI{22.9}{\%}$ of all $10^7\times10$ tracks have non-vanishing track probabilities for the best size estimates in the $R$ and $V$ band for either the thin disk model, Gaussian disk model or both disk models. 
Our python code to simulate the light curves, calculate $\chi^2$-values and produce the track library is available as well.\footnote{\url{https://github.com/sorgenfrei-c95/qsoMLsimcurves}}

\section{Results and Discussion}
\label{sec:results}

In Fig. \ref{fig:simcurves} we show the 25 best-fitting (in terms of track probability from Sect. \ref{sec:lcfit}) simulated light curves in $R$ and $V$ together with the observed $B-C$ data points, for the Gaussian (left panel) and thin disk model (right panel).
For both disk models it is noticeable that for almost all tracks the best-fitting simulated $V$ band model shows larger variations than the corresponding $R$ band model.
This result is consistent with the expected outwards-decreasing temperature profile from the theory of accretion disks (Sect. \ref{sec:intro}) and corresponds to a relative difference of the measured disk sizes in the $R$ and $V$ band (with smaller disks in the $V$ band) since larger sources show lower amplitude microlensing variations.
Note that only two of each 25 best-fitting tracks are identical for both disk models. Also, the shape of the light curves from the two models appears different, with more and sharper variations in the thin disk model hinting at subtle differences between the two models.
We stress that in the following, we consider all library tracks for our analysis.

\subsection{Size ratio and accretion disk temperature profile}
\label{sec:sizeratio}

For each of the $10^8$ tracks in our library we can determine the ratio of the sources size in the $R$ and $V$ band. This size ratio $q_{R/V}$ is directly related to the temperature profile (Eq. \ref{equ:tempprofil}) via an observed size dependence on the wavelength (Eq. \ref{equ:lambdasize}) of a Shakura-Sunyaev disk as discussed in Sect. \ref{sec:intro}. For the central filter wavelengths $\lambda_c(R)=\SI{6407}{\text{\AA}}$ and $\lambda_c(V)=\SI{5448}{\text{\AA}}$ \citep{Bessell_2005}, the expected theoretical size ratio therefore is $q_{R/V}^\text{theo.}\simeq1.241$ (i.e. a Shakura-Sunyaev disk is expected to appear $24.1\%$ larger in radius in the $R$ band compared to its size in the $V$ band).

As mentioned in Sect. \ref{sec:magpat}, we have chosen 80 different scale radius values non-logarithmically spaced, similar to \citet{Anguita_2008} and \citet{Eigenbrod_2008}. These were selected such that the resulting size ratios are as densely and close to evenly distributed in the ratio-space as possible. Therefore, for each track we computed the best fitting $R$ band over $V$ band half-light radii fraction $q_{R/V}:=r_{1/2}(R)/r_{1/2}(V)$ for the Gaussian and the thin disk model. We construct the histogram for $q_{R/V}$ by summing over the track probabilities $P(\chi^2)$ for a given size ratio bin (we use 16 ratio bins from $0.7$ to $2.2$ in steps of $0.1$). However, these bins are not exactly evenly sampled by the possible ratios. We correct for this by normalizing the probabilities by the number of source size combinations contributing to a particular bin. Overall, this is a minor correction since we selected the source sizes such that all their possible ratios are distributed as evenly as possible.

\begin{figure}
    \resizebox{\hsize}{!}{\includegraphics{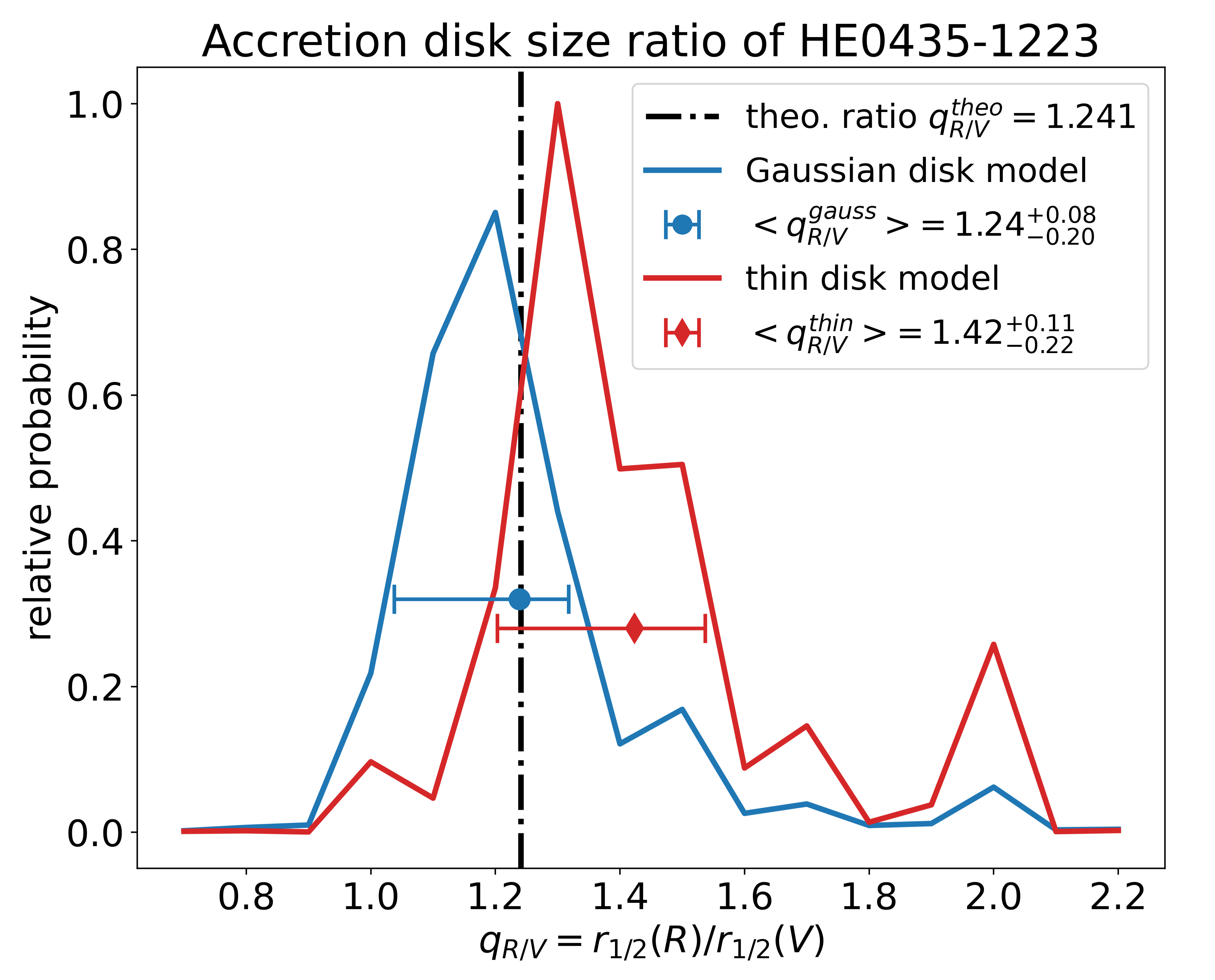}}
    \caption{Disk size ratio distribution. Shown are the probability distributions of the size ratio $q_{R/V}$ (ratio of $R$ and $V$ band size) for the Gaussian (blue) and the thin disk model (red) with corresponding mean values (blue circle for the Gaussian and red diamond for the thin disk model, each with $1\sigma$-uncertainties; note that their position along the ordinate is arbitrary), as well as the dash-dotted black line showing the theoretically expected value.}
    \label{fig:sizeratio}
\end{figure}

In Fig. \ref{fig:sizeratio}, we present the half-light radii ratio distributions $P(q_{R/V})$ for both disk models. The mean values with $1\sigma$-uncertainties are $\left<q_{R/V}\right>=1.24^{+0.08}_{-0.20}$ for the Gaussian disk model and $\left<q_{R/V}\right>=1.42^{+0.11}_{-0.22}$ for the thin disk model, both consistent with each other and the theoretically expected value. 
It can be noted, however, that the thin disk model prefers a somewhat larger ratio value, which corresponds to a shallower temperature profile according to Eq. \ref{equ:tempprofil}, with a negative exponent closer to $\beta=1/2$ (but still in agreement with $\beta=3/4$ from standard thin disk theory).\footnote{A direct conversion of the distribution of $q_{R/V}$ values to the often used negative temperature profile slope $\beta$ (see Eq. \ref{equ:tempprofil}) is problematic due to the pole at $q_{R/V}=1$. However, the inverse temperature profile slope $\zeta=1/\beta$, monotonically increasing as function of $q_{R/V}$, is accessible \citep[similar as in][]{Eigenbrod_2008}. We find $\left<\zeta\right>\simeq1.96^{+0.64}_{-0.83}$, again compatible with thin disk theory where $\zeta=4/3$, but somewhat shallower.} 
Such shallower temperature profiles have been measured and predicted as well as used to explain the typically observed size discrepancy of the disk appearing larger from microlensing studies as expected from their luminosity \citep{Poindexter_2008,Morgan_2010,Morgan_2018,Li_2019,Cornachione_2020a,Cornachione_2020c}.

Since our standard thin disk simulation concludes with a size ratio consistent with standard thin disk theory but with a tendency to a shallower $\beta=1/2$ temperature profile, we have run the same simulation again only changing the slope in Eq. \ref{equ:tempprofil} to $\beta=1/2$. Interestingly, all results almost perfectly agree with the $\beta=3/4$ simulation and we find $\left<q_{R/V}\right>=1.43^{+0.10}_{-0.23}$, close to the expected size ratio of $1.383$ for a thin $\beta=1/2$ disk as well as consistent with the ratio expected from standard theory. Therefore we cannot distinguish these two models from our analysis.

\subsection{Accretion disk size estimates}

The distribution of accretion disk sizes in units of Einstein radii can similarly be inferred from our simulation results; for each track in our library one needs to add the probability corresponding to the best fitting $R$ and $V$ band half-light radii to the appropriate histogram bins for both disk models.
We thus obtain four probability distributions $P(r_{1/2}|D)$ given our data $D$.
From these we find $\left<r_{1/2}\right>=0.91^{+2.14}_{-0.68}\,R_\text{E}$ for the Gaussian model in the $R$ band and $0.77^{+1.89}_{-0.56}\,R_\text{E}$ in $V$ band, while the thin disk model correspondingly gives $0.98^{+2.40}_{-0.73}\,R_\text{E}$ and $0.73^{+1.80}_{-0.56}\,R_\text{E}$.
We note that a quasar accretion disk of comparable size has recently been found by \citet{ForesToribio_2024} in the system SDSS J1004+4112. 

Moreover, since each track is associated with a velocity in $R_\text{E}/\SI{}{yr}$, we construct probability distributions $P(v|D)$ for both models, with mean track velocities of about $0.48$ and $\SI{0.33}{\textit{R}_E}/\SI{}{yr}$ for the Gaussian and the thin disk model, respectively. Combining both size and velocity information one can find the well-known linear source size--velocity degeneracy \citep[e.g.][]{Kochanek_2004}. 
We note that we also checked the distributions of the velocity direction and the compact fraction $\kappa_\star/\kappa$. We found only mild, but not significant, trends of smaller $\chi^2$ values towards velocities non-parallel to the shear direction and towards higher compact fraction maps.

One can convert the half-light radii $r_{1/2}$ (in $R_\text{E}$) to absolute sizes in $\SI{}{cm}$ by fixing the mean lens mass $\left<M\right>$ in Eq. \ref{equ:einsteinradius}. Alternatively, assumptions can be made about the velocities involved (source, lens plane, observer), which also translate into a mean mass of the compact objects. In fact, a probability distribution of the compact objects can be determined following the Bayesian method described in \citet[][especially Eqs. 13 to 18]{Kochanek_2004}. In brief, a prior effective source velocity probability $P(v_e)$ is constructed (with $v_e$ in $\SI{}{km/s}$) and convolved with the velocity likelihood function $P(v|D)$ from the simulation (with $v$ in $R_\text{E}/\SI{}{yr}$) to determine a lens mass distribution $P(\left<M\right>\!|D)$.

$P(v_e)$ then combines velocity contributions from the observer with respect to the CMB in the lens plane perpendicular to the direction to the quasar, Gaussian estimates for the peculiar velocities of lens galaxy and quasar, as well as the stellar velocity dispersion of the microlenses \citep[see][]{Kayser_1986,Kundic_1993,Kochanek_2004,Vernardos_2024} resulting in a broad probability density distribution with an average effective velocity of $521^{+212}_{-298}\,\SI{}{km/s}$. The velocity dispersion of the stars in the lens galaxy was measured by \citet{Courbin_2011} as $\sigma_\star=\SI{222}{km/s}$. 
We integrate the mass distribution $P(\left<M\right>\!|D)$ together with the half-light radii distribution $P(r_{1/2}(\left<M\right>)|D)$, where we used a constant mass prior from $0.1$ to $1.0\,M_{\odot}$ \citep{Kochanek_2004,Morgan_2018} and obtain a probability distribution for the absolute size of the accretion disk in $\SI{}{cm}$.

\begin{figure}
    \resizebox{\hsize}{!}{\includegraphics{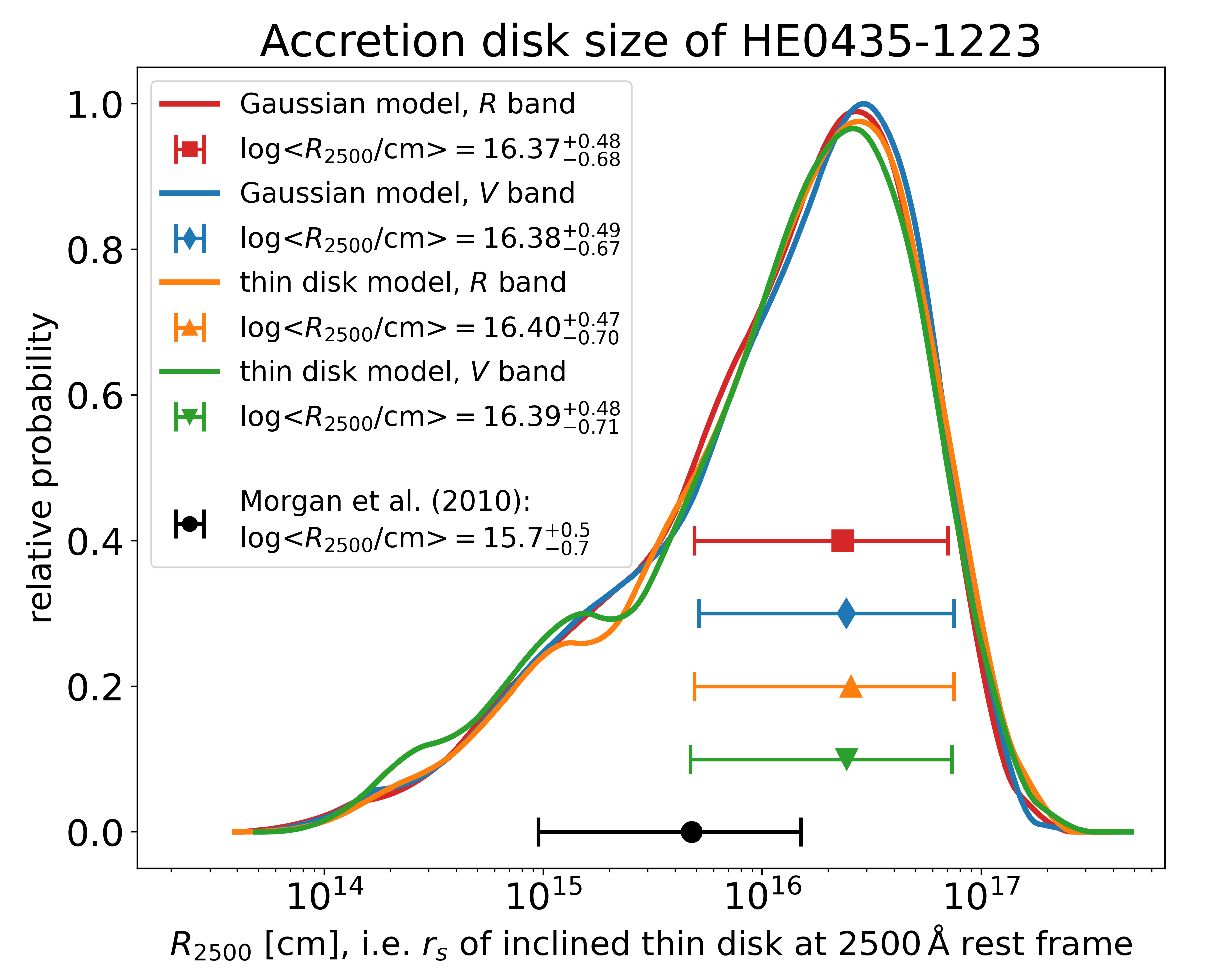}}
    \caption{Disk size estimates $R_{2500}$. Shown are the probability distributions for the accretion disk size of HE0435-1223 from our simulations using the Gaussian disk model with the $R$ band data (in red) and with the $V$ band data (in blue), as well as the thin disk model with $R$ (in orange) and with $V$ (in green). The (in colour) corresponding data points with error bars show the size expectation values with $1\sigma$-intervals. Similarly, the black data point shows the microlensing size result of HE0435-1223 found by \citet{Morgan_2010}. Note that the position of the data points along the ordinate is arbitrary.}
    \label{fig:disksizemorgan}
\end{figure}

To be able to compare the measured sizes to literature values, we convert the $r_{1/2}[\SI{}{cm}]$-values to the $R_{2500}[\SI{}{cm}]$ value from \citet{Morgan_2010}, where the size is expressed in terms of the thin disk scale parameter $r_s$ (see Eq. \ref{equ:brightnessprofil}, i.e. dividing the half-light radii by $2.44$, see Sect. \ref{sec:magpat}) at a UV-wavelength in the quasar rest frame of $\SI{2500}{\text{\AA}}$ (using Eq. \ref{equ:lambdasize}) of an inclined disk (assuming an average inclination angle of $i=60^\circ$ which increases the actual disk size by a factor of $\sqrt{2}$). Therefore we convert to
\begin{equation}
    R_{2500}=\frac{\sqrt{2}\,r_{1/2}}{2.44} \times\left[\frac{\SI{2500}{\text{\AA}}}{\lambda_c/(z_\text{s}+1)}\right]^{4/3}, 
\end{equation}
with the central filter wavelength $\lambda_c$ (see Sect. \ref{sec:sizeratio}) and the source redshift $z_\text{s}$ (see Sect. \ref{sec:intro}). 
The resulting size distributions $P(R_{2500}[\SI{}{cm}])$ for both models and bands are shown in Fig. \ref{fig:disksizemorgan}. We find $\log\left<R_{2500}/\SI{}{cm}\right>=16.37^{+0.48}_{-0.68}$ for the Gaussian disk in $R$ and $16.38^{+0.49}_{-0.67}$ in $V$ as well as $16.40^{+0.47}_{-0.70}$ for the thin disk in $R$ and $16.39^{+0.48}_{-0.72}$ in $V$, i.e. all around $R_{2500}\simeq\SI{2.4e16}{cm}$.

For comparison we note that \citet{Morgan_2010} found an accretion disk size of $\log\left<R_{2500}/\SI{}{cm}\right>=15.7^{+0.5}_{-0.7}$ for HE0435-1223 from a microlensing analysis, which is smaller but in agreement with our values. Our microlensing based disk sizes (as well as their value) are larger than their luminosity based estimate. In fact, \citet{Morgan_2010,Morgan_2018} show that such a discrepancy is found in many systems.

\subsection{Number of caustic crossings}
In the previous section we have derived average velocities of the quasar across the magnification patterns of $0.48$ or $\SI{0.33}{\textit{R}_E}/\SI{}{yr}$ for the Gaussian or think disk, respectively. During the approximately ten years of our observations, the quasar has thus moved (on average, approximately) three to five Einstein radii, which leads to the question of how many caustics the quasar has encountered or crossed in this time. We count here the caustic crossing of the disk centre, caustics that merely touch the outer part of the disk are not included. Consequently, the result will be a lower limit that is nevertheless indicative of the number of traversed caustic structures.

\begin{figure}
    \centering
    \resizebox{\hsize}{!}{
    \begin{subfigure}[c]{0.22\textwidth}
        \centering
        \includegraphics[height=0.2\textheight]{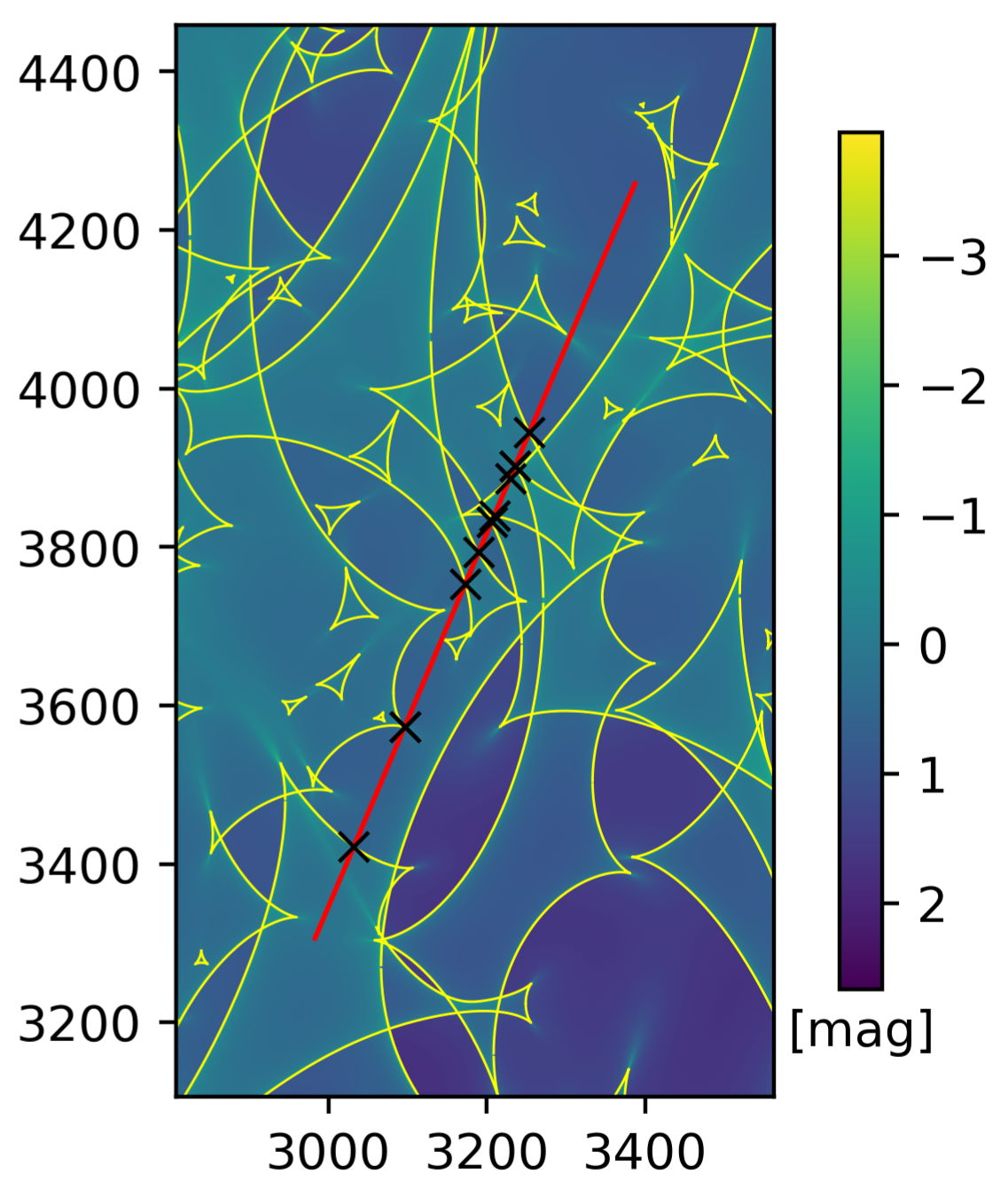}
        \caption{Example track}
        \label{fig:examplecrossing}
    \end{subfigure}
    \begin{subfigure}[c]{0.26\textwidth}
        \centering
        \includegraphics[height=0.2\textheight]{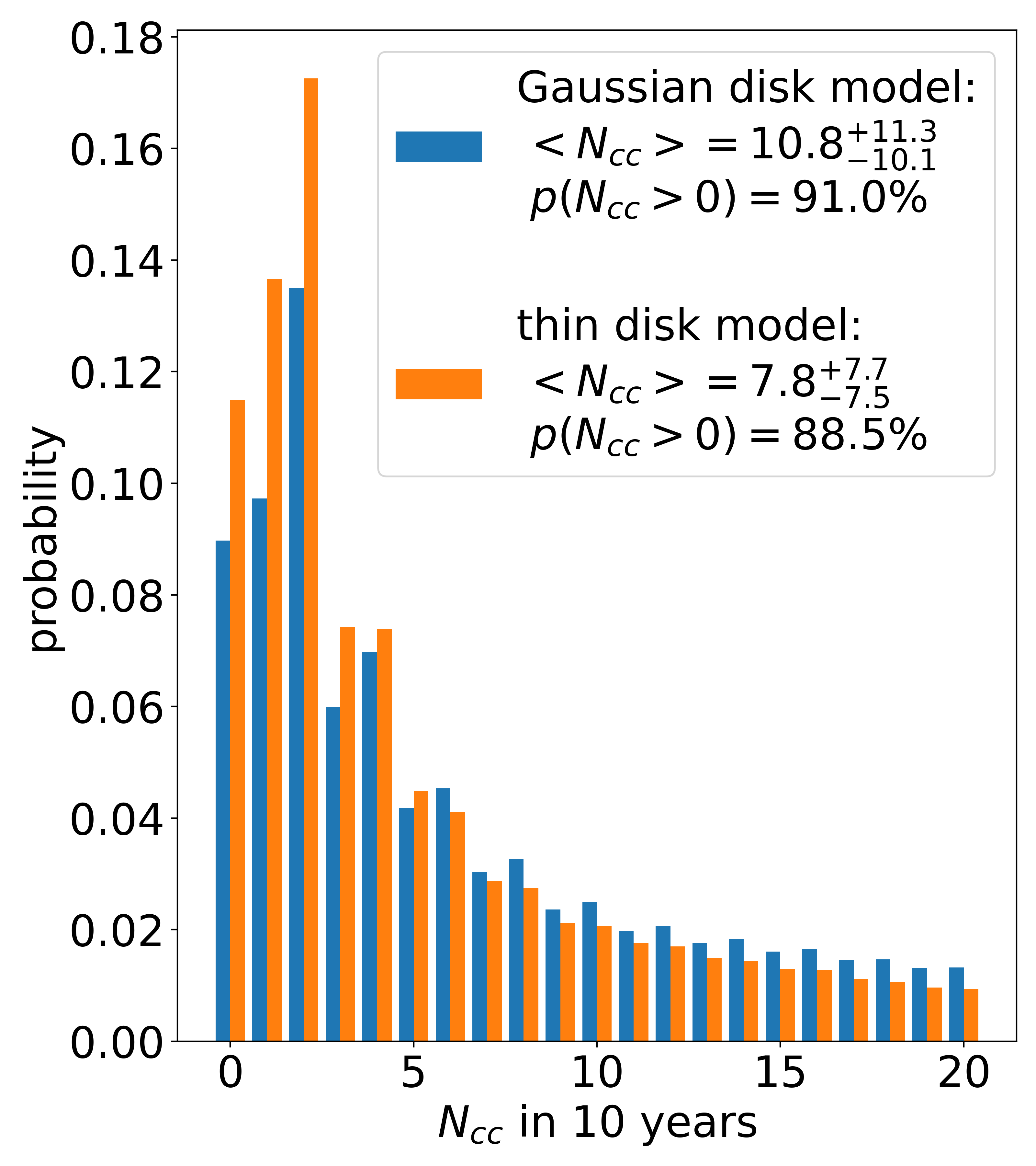}
        \caption{Number of caustic crossings}
        \label{fig:causticcrossings}
    \end{subfigure}}
    \caption{Number of caustic crossings $N_{cc}$. On the left, we show an example track (in red) on the $\kappa_\star/\kappa=1.0$ map (colour map in magnitudes, $x$- and $y$-axis in pixels) crossing $11$ (marked with black crosses) caustics (in yellow) during the $10$ simulated years. On the right, we show the resulting histogram of $N_{cc}$ (curtailed at $N_{cc}=20$) over all library tracks weighted by their probability.}
\end{figure}

We estimate the precise location of the caustics by using the complex parametrization of caustics by \citet{Witt_1990}, as implemented in \texttt{causticfinder-py}\footnote{\url{https://github.com/rschmidthd/causticfinder-py}}. We use the same positions and masses of the microlenses as used by \texttt{Teralens}\footnote{Note that the definition of the Einstein radius $R_\text{E}$ in \citet{Witt_1990} includes an additional factor of ${\sqrt{|1-\kappa_{\text{smooth}}|}}^{\,-1}$. The caustics thus determined agree perfectly with the magnification pattern calculated by \texttt{Teralens}, see Fig. \ref{fig:examplecrossing}.}. 
We count the number of caustic crossing $N_{cc}$ (in our case per $\SI{10}{yr}$, since this is the length the simulated tracks corresponds to) by calculating the number of intersections of each track with caustic lines for all tracks in our library with non-zero probability (see Fig. \ref{fig:examplecrossing} for an example track on the \texttt{Teralens} map together with the caustics calculated with \texttt{causticfinder-py}).
Note that not all caustic crossings lead to uniquely identifiable features in the light curves due to the (potentially small) separation along the track in time and the averaging effect from the source size.

Adding again the probabilities of each track in our library to the appropriate bins in a histogram of the number of caustic crossings, we obtain the probability distribution for $N_{cc}$ shown in Fig. \ref{fig:causticcrossings}. We find that the expected number of caustic crossings per \SI{10}{yr} $\left<N_{cc}\right>$ is $10.8^{+11.3}_{-10.1}$ and $7.8^{+7.7}_{-7.5}$ for the Gaussian and the thin disk model, respectively (with 16th and 84th percentiles as uncertainty). This essentially corresponds to around one caustic crossing per year, with a large spread.

The probability of having at least one caustic crossing in the \SI{10}{yr} interval can be integrated in both models to be about $90\%$. We stress again that the true number is even higher since we did not include the expanse of the disk. Given the size estimate from the last section, this makes caustic crossings in image B of HE0435-1223 very likely at all times. This could be especially interesting for studying this object in the X-ray regime \citep[e.g.][]{Guerras_2017} because not only is the X-ray source expected to be much smaller than the optical continuum region \citep{Pooley_2006,Zimmer_2011}, but also the emission from the smallest radii will be magnified most strongly in the X-ray spectra \citep[e.g.][]{Reynolds_2014,Mediavilla_2015,Chartas_2017}.

\section{Conclusion}
\label{sec:conclusion}
We have obtained $R$ and $V$ band observations of lensed quasars at Las Cumbres Observatory (LCO) for the past 10 years. From these we have determined light curves of the quasar images by applying DIA (together with PSF photometry and \textit{Gaia} proper motion data). In this study we have presented light curves of the four images of HE0435-1223 with 79 and 80 epochs in the two bands (Fig. \ref{fig:lightcurves}). We have determined difference curves (Fig. \ref{fig:diffcurves}) and find a strong microlensing signal in quasar image B.

We proceeded to analyse this signal by comparison with microlensing simulations (Sect. \ref{sec:MLsim}) using \texttt{Teralens} for the magnification patterns and the light curve simulating and fitting method developed by \citet{Kochanek_2004}. We applied our microlensing light curve simulations both to the Shakura-Sunyaev thin accretion disk model (Eq. \ref{equ:brightnessprofil}) and the Gaussian disk model (Eq. \ref{equ:gaussprofil}). Our analysis (Sect. \ref{sec:results}) shows that:
\begin{enumerate}
    \item The accretion disk is larger in the $R$ than in the $V$ band by a factor of $1.24^{+0.08}_{-0.20}$ in the Gaussian disk model and by a factor of $1.42^{+0.11}_{-0.22}$ in the thin disk model. This is our main result only made possible by the observations in two different photometric filters. These size ratios agree with the prediction from thin disk theory \citep{Shakura_1973}, from which a factor of $\sim\!1.241$ between the two bands is expected.
    \item The absolute size of the disk in terms of its half-light radius is around $0.7\,R_\text{E}$ to $1.0\,R_\text{E}$ (with an uncertainty of about $\SI{0.6}{dex}$) for the different models and bands, which we could all consistently convert to about $R_{2500}\simeq\SI{2.4e16}{cm}$ in terms of an inclined thin disks scale parameter at $\lambda_{\text{rest}}=\SI{2500}{\text{\AA}}$, i.e. larger but still in agreement with the value found by \citet{Morgan_2010}. In agreement with that study, our disk size measurement is thus also larger than predicted from their luminosity-based size estimate.
    \item Furthermore, we determined that on average (the centre of) image B crosses a caustic roughly once per year, explaining the long term fluctuations and the amplitude of the microlensing signal we find in the difference curves (Fig \ref{fig:diffcurves}).
\end{enumerate}

Our results are consistent between the two disk models, as was suggested by the results from \citet{Mortonson_2005}. We also see small differences, such as the shape of the light curves (Fig. \ref{fig:simcurves}) and the marginally different size ratios (Fig. \ref{fig:sizeratio}). Further studies testing multiple models could be used to find out whether such hints at deviations from the standard Gaussian brightness profile can be found in other systems as well. Also, since different studies find temperature profile slopes deviating from thin disk theory \citep[e.g.][which concludes that shallower slopes are favoured incorporating results from multiple studies]{Cornachione_2020c}, including models with shallower slopes such as $\beta=1/2$ (which is also close to the value we find for the thin disk model and therefore tested as well as described in Sect. \ref{sec:sizeratio}) could be revealing. This is especially the case, since such shallower temperature profiles can be used to explain the typically found accretion disk size discrepancy between large microlensing and smaller luminosity based estimates \citep{Morgan_2010}.

We have updated our LCO light curve data for the three quasars in \citet{Sorgenfrei_2024} and are working on more lensed quasars to find suitable chromatic microlensing events to analyse. In the future, multi-band data from the Legacy Survey of Space and Time (LSST) of the \textit{Vera C. Rubin Observatory} will produce huge amounts of data ideal for chromatic quasar microlensing studies \citep{Ivezic_2019}, helping to constrain the structure of quasar accretion disks further.

\section*{Data availability}
The $R$ and $V$ band light curves of HE0435-1223 from Fig. \ref{fig:lightcurves} are available in electronic form at the CDS. 
The updated light curves of the four quasars described in Sect. \ref{sec:intro} are available at \cite{GAVO_2025}. 

\begin{acknowledgements}
This work makes use of observations from the Las Cumbres Observatory global telescope network.
The authors acknowledge support by the High Performance and Cloud Computing Group at the Zentrum für Datenverarbeitung of the University of Tübingen, the state of Baden-Württemberg through bwHPC and the German Research Foundation (DFG) through grant no INST 37/935-1 FUGG.
This work has made use of data from the European Space Agency (ESA) mission {\it Gaia} (\url{https://www.cosmos.esa.int/gaia}), processed by the {\it Gaia} Data Processing and Analysis Consortium (DPAC, \url{https://www.cosmos.esa.int/web/gaia/dpac/consortium}). Funding for the DPAC has been provided by national institutions, in particular the institutions participating in the {\it Gaia} Multilateral Agreement.
This work made use of Astropy (\url{http://www.astropy.org}).
C.S. acknowledges support from the International Max Planck Research School for Astronomy and Cosmic Physics at the University of Heidelberg.
We thank Aksel Alpay for support with \texttt{Teralens}, Markus Demleitner for making the data available at GAVO, as well as Markus Hundertmark, Yiannis Tsapras, Zofia Kaczmarek and David Kuhlbrodt for helpful discussions.
Finally, we thank the anonymous referee for the helpful report.
\end{acknowledgements}

\bibliographystyle{aa}
\bibliography{BibTeX_CS}

\end{document}